# Broadband on-Chip Optical Non-reciprocity Using Phase Modulators

Christophe Galland[1], Ran Ding[1], Nicholas C. Harris[1], Tom Baehr-Jones[1], Michael Hochberg[1,2,3]

*[1]Department of Electrical & Computer Engineering, University of Delaware, Newark DE 19716, USA*
*[2]Department of Electrical & Computer Engineering, National University of Singapore, Singapore 117576*
*[3]Institute of Microelectronics, A*STAR (Agency for Science, Technology and Research), 11 Science Park Road, Singapore Science Park II, Singapore 117685*

**Abstract:** Breaking the reciprocity of light propagation in a nanoscale photonic integrated circuit (PIC) is a topic of intense research, fostered by the promises of this technology in areas ranging from experimental research in classical and quantum optics to high-rate telecommunications and data interconnects. In particular, silicon PICs fabricated in processes compatible with the existing complementary metal-oxide-semiconductor (CMOS) infrastructure have attracted remarkable attention. However, a practical solution for integrating optical isolators and circulators within the current CMOS technology remains elusive. Here, we introduce a new non-reciprocal photonic circuit operating with standard single-mode waveguides or fibers. Our design exploits a time-dependent index modulation obtained with conventional phase modulators such as the one widely available in silicon photonics platforms. Because it is based on fully balanced interferometers and does not involve resonant structures, our scheme is also intrinsically broadband. Using realistic parameters we calculate an extinction ratio superior to 20 dB and insertion loss below -3 dB.

## 1. Introduction

Materials typically used to guide and manipulate light in optical fibers and photonic integrated circuits (PIC) have symmetric permittivity and permeability tensors and in the linear regime light propagation satisfies the Lorentz reciprocity theorem [1]. One way to break reciprocity is to operate in the non-linear regime and harvest bi-stability [2]. A more versatile approach is to guide light through a material exhibiting strong magneto-optical effect, which causes a Faraday rotation of the polarization dependent on the propagation direction [3-5]. In practice, this effect is used in commercial optical isolators and circulators, which are essential components in today's fiber-optics systems. For example, protecting lasers from back-scattered light is critical to ensure their integrity and stability. Yet, whereas fiber-optics isolators and circulators relying on the magneto-optic effect have become widely available, a practical solution for PICs has yet to be demonstrated. Obtaining non-reciprocity with the help of magneto-optic materials by constructing a hybrid chip [6, 7] suffers from much increased fabrication complexity. This approach suppresses most benefits of silicon PICs, which rely on their entire fabrication in a CMOS-compatible process that ensures scalability, yield and low cost.

Recently, the first electrically driven non-reciprocal device in a silicon PIC was reported [8], based on the concept of interband photonic transitions developed earlier by Yu and Fan [9, 10] (an idea similar to the one already used for non-reciprocal mode-conversion in optical fibers [11]). In this device, a traveling-wave radio-frequency (RF) signal induced a time-varying and spatially non-homogeneous modulation of the refractive index in a silicon waveguide specially engineered to support two TE modes with opposite symmetries and different propagation constants and wavevectors. While the concept is elegant and no hybrid technology is needed, the implementation is so far prohibitively complicated, and the fabricated device exhibited >70 dB insertion loss [8].

It is therefore intriguing to ask whether a simpler scheme, relying only on existing components that can be readily fabricated in a CMOS process, can be exploited to obtain non-reciprocal light propagation in a monolithic silicon PIC. Others have shown interesting progress in this direction by employing two "tandem" phase modulators to imprint a non-reciprocal frequency shift [12]. Two limitations of this scheme are its intrinsic narrow-band operation and its modest extinction ratio of 10.8 dB. Passive resonant structures can be employed to enhance intrinsic silicon non-linearities [2], but this approach also suffers from narrow optical bandwidth and the performance intrinsically depends on the input light power. Finally, non-reciprocal light modulation can be achieved in traveling-wave modulators, at the cost of very long devices operating at very high-speed, and with only modest extinction ratio [13].

In this Letter, we propose and analyze a novel non-reciprocal device utilizing balanced interferometers and multi-stage phase modulation at a frequency of a few GHz or less. Our scheme is broadband by design and can achieve an extinction ratio of better than -20 dB with insertion loss smaller than -3 dB, while being readily manufacturable in a standard CMOS photonic process. Our proposal relies on phase modulators separated by passive waveguide sections and driven by delayed versions of the same RF signal. The engineered RF and optical delay between modulator sections ensures non-reciprocity [10], and thus our scheme does not require very long traveling-wave modulators nor very high operation frequency, as was the case in previous publications [8, 13, 14], reducing simultaneously the technological complexity, the insertion loss and the achievable footprint [15]. In a single pass, our scheme acts as a non-reciprocal modulator that can be used to encode return-to-zero (RZ) symbols. Total extinction of counter-propagating light is achieved independently of the driving signal shape and amplitude. Cascading two devices, we achieve isolation with predicted performances comparable to what is obtainable with fiber-based components, while the transmitted signal's phase and amplitude remain non-modulated. We emphasize that none of our arguments relies on mode conversion [16] and that the device operates on single-mode waveguides, as required for the waveguides used in the dominant silicon-on-insulator PIC technology [17]. To our knowledge, these are the first practical designs proposed for on-chip broadband optical non-reciprocity and they should open new avenues in the development and applications of integrated photonics.

## 2. Single-stage non-reciprocal modulator

*2.1 System design*

The design schematized in Fig. 1a. consists of two Mach-Zehnder modulators (MZMs) driven by the same RF signal and separated by an optical delay line inducing a quarter-period retardation in the signal driving MZM *a* (left-hand side) with respect to MZM *b* (right-hand side). (An alternative and equivalent design based on a four-stage modulator is presented in section 5.3.) We note that tunable RF delay lines capable of 100ps delay with little distortion on 10Gb/s data stream can readily be implemented in a standard CMOS circuit [18] that could be wire- or flip-chip-bonded to the photonic chip. We consider a periodic function $F(t)$: $F(t) = F(t+T)$; with period $T = 1/f$; also satisfying $F(t\pm T/2)=-F(t)$, and normalized to have peak-to-peak amplitude ±1. Examples of such functions are sine and cosine waves, as well as square waves with 50% duty cycle. An MZM is implemented by modulating the optical phase in each arm of the Mach-Zehnder interferometer. For chirp-free operation, the phase modulators are driven in push-pull mode: the upper arm experiences a phase shift $\varphi(t)$ while the lower arm is driven symmetrically with a phase shift -$\varphi(t)$ with respect to a constant offset. In the laboratory time frame of reference, the optical phase modulations in MZM *a* and *b* can be written $\varphi_a(t)=\gamma(1\pm F(t-T/4))$ and $\varphi_b(t)=\gamma(1\pm F(t))$, respectively, with $\gamma$ the effective phase modulation amplitude (in radians) and +/- for the upper/lower arm. By choosing the waveguide length between the two MZMs to be $L_{opt} = T/4\ c/n_g$ ($n_g$ is the group index and $c$ the speed of light in vacuum) light incoming from the right (ports $b_1$, $b_2$) travels in phase with the RF signal and therefore experiences twice the same modulation in MZMs *b* and *a*. On the contrary, for light incoming from the left (port $a_1$, $a_2$), the modulation functions in MZMs *a* and *b* exhibit a $\pi$ phase shift (i.e. have opposite signs). In this configuration, non-reciprocity is ensured by the presence of two phase-shifted periodic signals. Since high-visibility interference requires the optical path difference between the two arms of the delay line to be adjustable (see below), resistive heaters shall be used to finely tune their relative index [19]. In the following simulations we will take $f$ = 4 GHz, which for a group index of $n_g$ = 4.2 [20] leads to $L_{opt}$ = 4.46 mm.

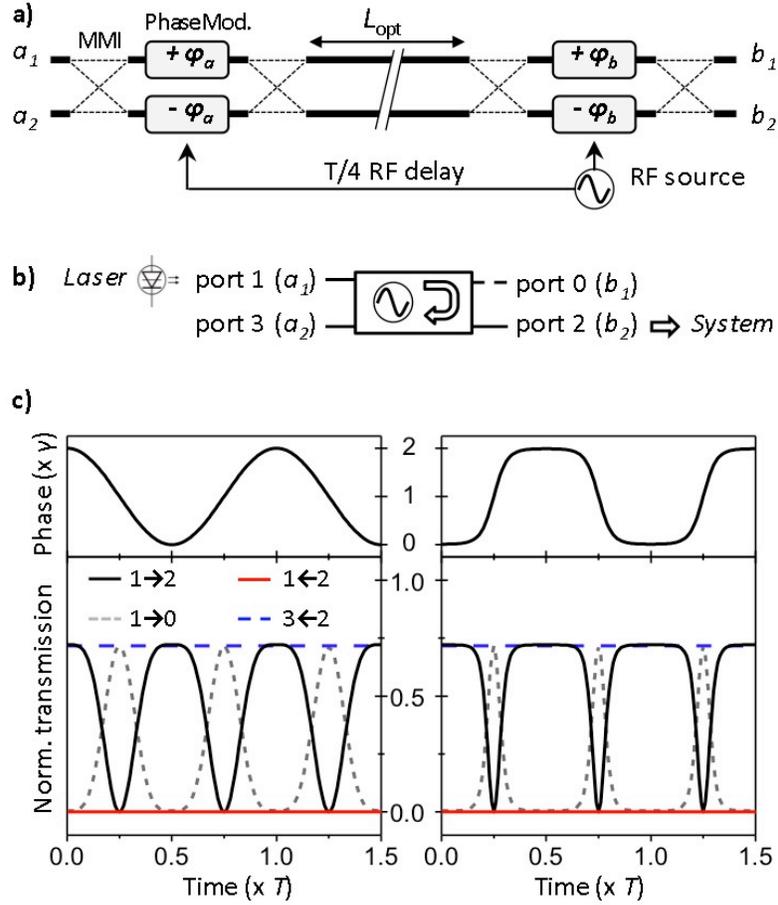

**Figure 1.** a) Schematic of the proposed design. MMI: multimode interferometer. The waveguide length of the optical delay is $L_{opt}= T/4\ c/n_g$. b) Conceptualization of the device as a non-reciprocal modulator. c) Computed behavior for a cosine (left panels) and a bandwidth-limited square wave (right panels). The waveforms are shown in the upper panels. In the lower panels, we plot the transmission coefficients from port 1 to 2 (solid black line), 2 to 1 (solid red line), 1 to 0 (dashed grey line) and 2 to 3 (dashed blue line). We note that the device is symmetric under the simultaneous permutation 1↔3 and 0↔2. For the simulations we used the following parameters: MMI loss = -0.1 dB [21, 22]; waveguide loss = -0.3 dB/cm [23]; total waveguide length = 8 mm; dynamic loss = -2 dB / $\pi$ phase shift (see section 5.2).

*2.2 Transfer Matrix calculations*

To give an intuitive understanding of the functioning of our device, we present first a simplified calculation of its idealized transfer matrix. Fully realistic simulations are used next to compute the accurate behavior (see section 5.1 for details). Within the coupled-mode formalism, the relationship between complex field amplitudes $a_1$, $a_2$ and $b_1$, $b_2$ at the input and output of a four-port device, written in vector notation $\begin{pmatrix} a_1 \\ a_2 \end{pmatrix}$ ; $\begin{pmatrix} b_1 \\ b_2 \end{pmatrix}$, is expressed through a 2x2 matrix. The transfer matrix of an ideal 50/50 beam splitter (directional coupler or multimode interferometer) writes:

$$BS = \frac{1}{\sqrt{2}} \begin{pmatrix} 1 & i \\ i & 1 \end{pmatrix}$$

And that of an ideal MZM driven in push-pull mode:

$$M(\varphi) = \frac{1}{2}\begin{pmatrix} \exp(i\varphi) - \exp(-i\varphi) & i(\exp(i\varphi) + \exp(-i\varphi)) \\ i(\exp(i\varphi) + \exp(-i\varphi)) & -\exp(i\varphi) + \exp(-i\varphi) \end{pmatrix}$$

We can thus write the transfer matrix of the circuit in Fig. 1a for light propagating from *a* to *b* (left to right):

$$\vec{T} = \frac{1}{4}\begin{pmatrix} 1 & i \\ i & 1 \end{pmatrix} \times \begin{pmatrix} \exp(i\varphi_b) & 0 \\ 0 & \exp(-i\varphi_b) \end{pmatrix} \times \begin{pmatrix} 1 & i \\ i & 1 \end{pmatrix}^2 \times \begin{pmatrix} \exp(i\vec{\varphi}_a) & 0 \\ 0 & \exp(-i\vec{\varphi}_a) \end{pmatrix} \times \begin{pmatrix} 1 & i \\ i & 1 \end{pmatrix}$$

Where we have defined $\vec{\varphi}_a(t) = \gamma F(t - \pi/\Omega) = -\varphi_b(t)$ the time-dependent optical phase shift experienced in MZM *a* with reference to the one experienced in MZM *b*. Similarly, for light propagating from *b* to *a* (right to left), the transfer matrix of the complete device is:

$$\bar{T} = \frac{1}{4}\begin{pmatrix} 1 & i \\ i & 1 \end{pmatrix} \times \begin{pmatrix} \exp(i\bar{\varphi}_a) & 0 \\ 0 & \exp(-i\bar{\varphi}_a) \end{pmatrix} \times \begin{pmatrix} 1 & i \\ i & 1 \end{pmatrix}^2 \times \begin{pmatrix} \exp(i\varphi_b) & 0 \\ 0 & \exp(-i\varphi_b) \end{pmatrix} \times \begin{pmatrix} 1 & i \\ i & 1 \end{pmatrix}$$

with the time-dependent phase-shift experienced by the light at MZM *a* now being $\bar{\varphi}_a(t) = \gamma F(t) = \varphi_b(t)$. Note that we have factored out the constant phase shift $+\gamma$ accumulated in each arm. Evaluating these expressions yields:

$$\vec{T} = \begin{pmatrix} -\cos(2\gamma F(t)) & -\sin(2\gamma F(t)) \\ \sin(2\gamma F(t)) & -\cos(2\gamma F(t)) \end{pmatrix} \text{ and } \bar{T} = \begin{pmatrix} -1 & 0 \\ 0 & -1 \end{pmatrix}$$

As anticipated, the transfer matrix is non-reciprocal. Figure 1b presents how our system can be conceptualized as an optical circulator by assigning ports 1, 2 and 3 to $a_1$, $b_2$ and $a_2$, respectively (port 0 = $b_1$ can be used for monitoring or can be terminated by a taper to avoid reflection). From the above expression for $\bar{T}$ it can be seen that port 1 is perfectly isolated from light incident in port 2, independently of the precise shape and amplitude of the RF signal $F(t)$, as long as the condition $L_{opt} = T/4 \, c/n_g$ is fulfilled.

*2.3 Numerical simulations*

We show the simulated behavior of a realistic device in Fig. 1c, for a phase modulation amplitude $\gamma = \pi/4$ ($\pi/2$ peak-to-peak) and two different waveforms (a cosine and a bandwidth-limited square wave). In the calculations we included physically relevant effects such as waveguide loss (-0.3 dB/cm) [23] and beam splitter insertion loss (-0.1 dB) [21, 22], and the dynamic loss intrinsically linked to phase modulation using free-carrier dispersion effect [24] (see section in section 5.2). Even in the presence of losses, and independent of the signal shape, transmission from port 2 to 1 is exactly zero, demonstrating robust isolation. In contrast, transmission from port 1 to 2 is non-zero, with a maximal value of ~ 0.74 (or -1.3 dB) and a time-averaged value of –3.27 dB for a cosine signal. Improved averaged transmission is achieved by driving the modulators with a square-wave. For example, assuming a modulation bandwidth of 5*f* (Fig. 1c, right panel), time-averaged insertion loss decreases to –2.17 dB.

This first result is by itself remarkable and potentially useful in real systems. For example, the output at port 2 can be directly used as a clock signal, or as an information carrier in return-to-zero encoding schemes. The device can even be used to simultaneously perform the encoding by modulating the amplitude $\gamma$. Our system therefore integrates a return-to-zero modulator and a high extinction isolator into a single compact device, making use of only conventional components. We emphasize that the modulation frequency can be chosen arbitrarily, as low as permitted by waveguide propagation losses in the delay line and desired footprint, and as fast as permitted by the modulators and drivers bandwidths.

## 3. Double-stage isolator

*3.1 System design*

We now consider the device used in reversed direction (or equivalently with an opposite sign of the RF delay). As seen in Fig. 1c, we have non-modulated transmission from port 2 to 3, independently of the driving signal parameters, yet there are spikes of transmission from port 3 to 2, preventing complete isolation. We therefore propose to cascade two identical devices and drive them with the same RF source, with a quarter-period RF delay between them as shown in Fig. 2a. The two devices should be positioned immediately next to each other so that negligible optical delay is introduced between them. In this scheme, light passing through the transmission spikes of the first device is always rejected by the second one.

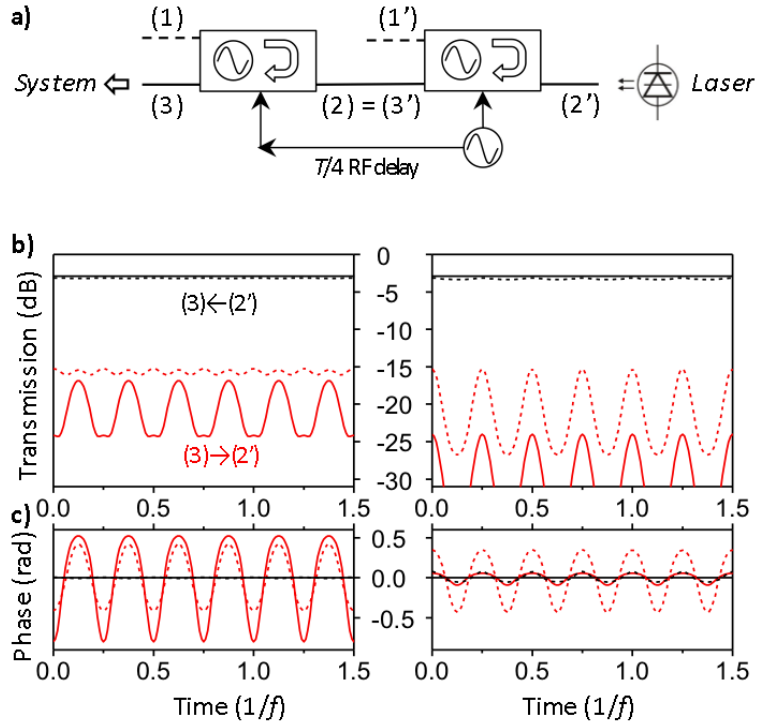

**Figure 2.** a) Schematics of the cascaded configuration used to achieve isolation without modulation of the transmitted light. b-c) Computed amplitude (b) and phase (c) transmission coefficients (amplitude is plotted in dB; phase variation is given in radian relative to an arbitrary reference). The left and right panels correspond to cosine and smoothed square-wave RF waveforms as shown in Fig. 1c. Black lines are for propagation from port 2' to 3 and red line from port 3 to 2'. Solid lines are obtained for perfectly balanced devices, while for dashed lines an arm imbalance of $\pi/10$ rad (optical phase) is introduced in each device.

*3.2 Numerical simulations*

The simulations in Figs. 2b-c demonstrate that this configuration indeed achieves non-modulated transmission of right-to-left propagating light, both in amplitude (Fig. 2b) and phase (Fig. 2c), with insertion loss of -2.9 dB. Extinction is reduced compared to Fig. 1c but still better -20 dB in the case of square-wave modulation and perfect arm balancing (-14 dB for cosine modulation). This figure can be further improved by increasing the bandwidth-to-frequency ratio. When the optical paths in the arms of each delay line are not perfectly balanced, the performance are slightly degraded, as shown by the dashed lines in Figs. 2b-c for an optical phase mismatch of $\pi/10$.

To estimate the impact of relevant parameters and imperfections on the system, we study two figures of merit: the insertion loss (IL), defined as the time-averaged transmission in the passing direction, and the extinction ratio (ER), equal to the peak transmission value in "blocking" direction divided by the IL. In Fig. 3a, we show the dependence of IL and ER on the modulation amplitude for the single-pass and the cascaded configurations; the optimal modulation amplitude in both cases is close to $\gamma = \pi/4$, which yields the lowest IL for single-pass configuration and highest ER for cascaded configuration.

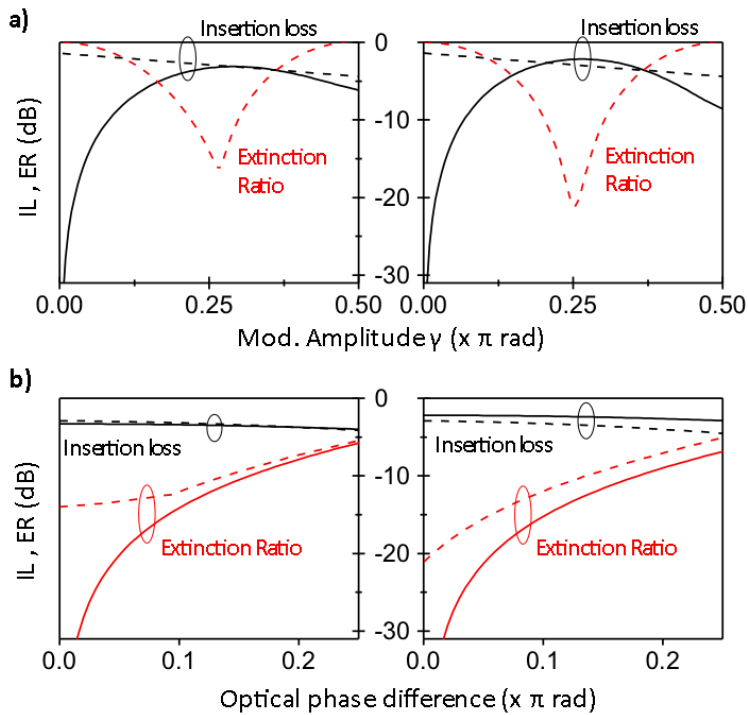

**Figure 3**. a) Insertion loss (IL, black curves) and extinction ratio (ER, red curves) as a function of the modulation amplitude $\gamma$ (in units of $\pi$ rad), for the single-pass configuration of Fig. 1a (solid lines) and the cascaded setting of Fig. 2a (dashed lines). b) Same performances plotted as a function of the optical phase imbalance in the delay line of each device, for a fixed modulation amplitude $\gamma = \pi/4$. Color scheme as in a).

Figure 3 also reports the sensitivity of the performances on the relative phase difference accumulated in the two arms of the delay line (modulo $2\pi$). Given the thermo-optic coefficient of silicon $dn_{Si}/dT = 1.9 \times 10^{-4}$ K$^{-1}$ around 1.55µm, we calculate the temperature-dependent phase shift (in rad) per unit length of waveguide to be less than $\pi/4$ mm$^{-1}$K$^{-1}$. Controlling the phase difference within $\pi/10$ can thus be achieved by tuning the temperature over a 0.4 mm section with 1 K accuracy, well within reach of existing technology [25]. Yet some feedback circuit will likely be needed to ensure stable operation.

## 4. Conclusion

To conclude, we estimate the expected optical bandwidth. As the system relies on a series of fully balanced interferometers, it is by design wavelength insensitive. Yet, several second-order effects may limit the actual operating wavelength range. Directional couplers usually have limited bandwidth, so we opt for MMIs [21, 26] (see Fig. 1a), for which uniform splitting ratio over 94 nm have been demonstrated [27]. Second, due to group velocity dispersion the optical delay between MZM *a* and *b* is actually wavelength dependent. Using the dispersion measured in [20] and a waveguide length of 4.7 mm, we estimate a delay variation of less than 1 ps over more than 100 nm bandwidth around 1550 nm. This is much smaller than the modulation period (250 ps here) and has therefore negligible impact on performance (the ratio ~1/250 is independent of the particular modulation frequency). This variation could be further reduced by tailoring the dispersion [28]. The limiting factor may eventually come from the wavelength dependence of the plasma dispersion effect [24], but this too could be easily compensated by tuning the modulation amplitude $\gamma$.

**Acknowledgement**

The authors would like to thank Gernot Pomrenke, of the Air Force Office of Scientific Research, for his support under the OPSIS (FA9550-10-1-0439) PECASE (FA9550-10-1-0053) and STTR (FA9550-12-C-0079) programs, and would like to thank Mario Paniccia and Justin Rattner, of Intel, for their support of the OpSIS program. Professor Hochberg would like to acknowledge support from the Singapore Ministry of Education ACRF grant R-263-000-A09-133.

## 5. Appendices

*5.1 Transfer Matrix Calculations*

In this section we details the calculations of the transfer matrices used to produce the simulations shown in the paper. Within the coupled-mode formalism the complex field amplitudes $a_1$, $a_2$ and $b_1$, $b_2$ at the input and output of a four-port device are written as vectors $\begin{pmatrix} a_1 \\ a_2 \end{pmatrix}; \begin{pmatrix} b_1 \\ b_2 \end{pmatrix}$ and the input-output relationship takes the form of a complex 2x2 matrix. We note that all following calculations are up to a common phase factor to both arms of the circuit. We list below the elementary components into which our non-reciprocal devices (Fig. 1a in main text) can be decomposed, and derive their respective transfer matrices.

- The transfer matrix of a beam splitter (BS) – either a directional coupler or a multimode interferometer – having splitting ratio $r$ and insertion loss $k$ (in dB) writes:

$$BS(r,k) = \sqrt{10^{-k/10}} \begin{pmatrix} \sqrt{1-r} & i\sqrt{r} \\ i\sqrt{r} & \sqrt{1-r} \end{pmatrix}$$

- In a Mach-Zehnder modulator (MZM) driven in push-pull mode, the upper arm experiences a phase shift $\varphi(t)$ while the lower arm is driven symmetrically with a phase shift $-\varphi(t)$. This is obtained by applying an offset bias and modulating each arm with opposite voltage signs around this offset. We also introduce an arm imbalance quantified by a phase delay $\partial\varphi$ in the upper arm relative to the lower arm. We furthermore account for dynamic losses caused by free carrier absorption, which always accompanies phase modulation by plasma dispersion effect (see next section). They are characterized by an excess loss per $\pi$ phase shift of ß (in dB). Under increasing free carrier concentration, the refractive index (and therefore the optical phase delay) decreases, so that, with reference to the intrinsic waveguide delay, the phase modulation term writes $\exp\{-i\cdot(-\gamma\{1+F(t)\})\} = \exp\{i\cdot\gamma\{1+F(t)\}\}$. The dynamic loss increases under carrier injection and is thus expressed as $\exp\{-\alpha\cdot\gamma\{1+F(t)\}\}$; where we have defined $\alpha = \dfrac{\beta\ln(10)}{\pi\cdot 20}$. In the opposite arm the factor $1+F(t)$ is replaced by $1-F(t)$ to simulate push-pull operation. The transfer matrix for the modulator section in an MZM is then:

$$MZ(\gamma,\alpha,\delta\varphi,F(t)) = \begin{pmatrix} \exp\{-i\partial\varphi + (-\alpha+i)\cdot\gamma\cdot(1+F(t))\} & 0 \\ 0 & \exp\{(-\alpha+i)\cdot\gamma\cdot(1-F(t))\} \end{pmatrix}$$

These two general matrices suffice to express all the parts of the circuits considered in our paper. In particular, waveguide loss in any section can be introduced with $BS(r=0, k>0)$, while an imbalance between the upper and lower arm of the delay line is simulated by inserting $MZ(\gamma=0, \alpha=0, \delta\varphi, F(t)=0)$

Our model can be used to study all relevant optical effects in the linear regime, as well as arbitrary RF signals (including bandwidth limitation, by appropriate choice of the function *F(t)*), and of course the impact of non-ideal physical implementation such as MZ arm imbalance, asymmetric splitting ratios, etc.

*5.2 Dynamic loss and free carrier modulation*

Here we detail how we obtain the value for the dynamic modulation loss that we use in the simulations. High-speed (>20 GHz) phase modulation in Si is achieved through the "plasma dispersion" effect, i.e. the change in the complex dielectric function due to a change in free carrier concentration. Soref and Bennett derived the magnitude of this effect in 1987, using a combination of previous experimental data:

$$\Delta n = -8.8 \times 10^{-22} \Delta N_e - 8.5 \times 10^{-18} \Delta N_h^{0.8}$$

$$\Delta \alpha = 8.5 \times 10^{-18} \Delta N_e + 6.0 \times 10^{-18} \Delta N_h$$

Here, $\Delta n$ is the change in refractive index, $\Delta \alpha$ is the change in absorption [in cm$^{-1}$], and $\Delta N_e ; \Delta N_h$ the free electron and hole concentrations [in cm$^{-3}$], respectively.

Using these relations we can derive the excess loss [in dB] caused by free carrier absorption for a change of index $\Delta n$ over a propagation length *L*, when only free electrons are considered:

$$\beta_e \text{ [dB]} \approx -4.2 \times 10^4 \Delta n \times L[\text{cm}]$$

And similarly for free holes only:

$$\beta_h \text{ [dB]} \approx -5.7 \times 10^4 \Delta n^{1.25} \times L[\text{cm}]$$

Accumulating a π optical phase delay corresponds to the condition $\Delta n \times L = \lambda / 2$ with $\lambda = 1.55 \times 10^{-4}$ cm the vacuum wavelength. This leads to the expression for the excess loss *intrinsically* related to a π phase shift in a device of length *L*:

$$\beta_{\pi,e} \approx -3.25 \text{ dB}$$

$$\beta_{\pi,h} \approx -0.5 \times L[\text{cm}]^{-0.25} \text{ dB}$$

Since both electrons and holes contribute to the plasma dispersion effect in depletion-based PN modulators or injection-based PIN modulators, we use and average value of $\beta_\pi \equiv \beta \approx -2$ dB in the simulations. An interesting question is whether hole-only devices (based on capacitor structures with p-doped waveguides) could be fabricated to lower the dynamic loss in phase modulators, especially in the limit of very long, lightly doped devices.

*5.3 Alternative Proposal*

We propose a second scheme pictured in Fig 4. Here, a single MZI is built with four phase modulators in each arm, pair-wise driven in push-pull mode by the signal *F(t)*. Again, the electric signal is delayed from right to left by a quarter-period between each successive pair of modulators, and an optical path length $L_{opt}$ = $T/4 \, c/n_g$ is introduced to ensure that light launched from the right keeps a fixed relationship with the phase of the electric drive, thus experiencing a total optical phase modulation $\overleftarrow{\varphi}(t) = 4\gamma(1 \pm F(t))$. Light propagating from left to right sees a π retardation in the electric signal between each successive modulator, thus accumulating a zero net phase shift relative to the other arm: $\Delta \vec{\varphi}(t) = 2\gamma(F(t) + F(t - \pi / \Omega)) = 0$. The transfer matrices for this configuration are therefore:

$$\vec{S} = i \begin{pmatrix} \sin(4\gamma F(t)) & \cos(4\gamma F(t)) \\ \cos(4\gamma F(t)) & -\sin(4\gamma F(t)) \end{pmatrix} \text{ and } \overleftarrow{S} = i \begin{pmatrix} 0 & 1 \\ 1 & 0 \end{pmatrix}$$

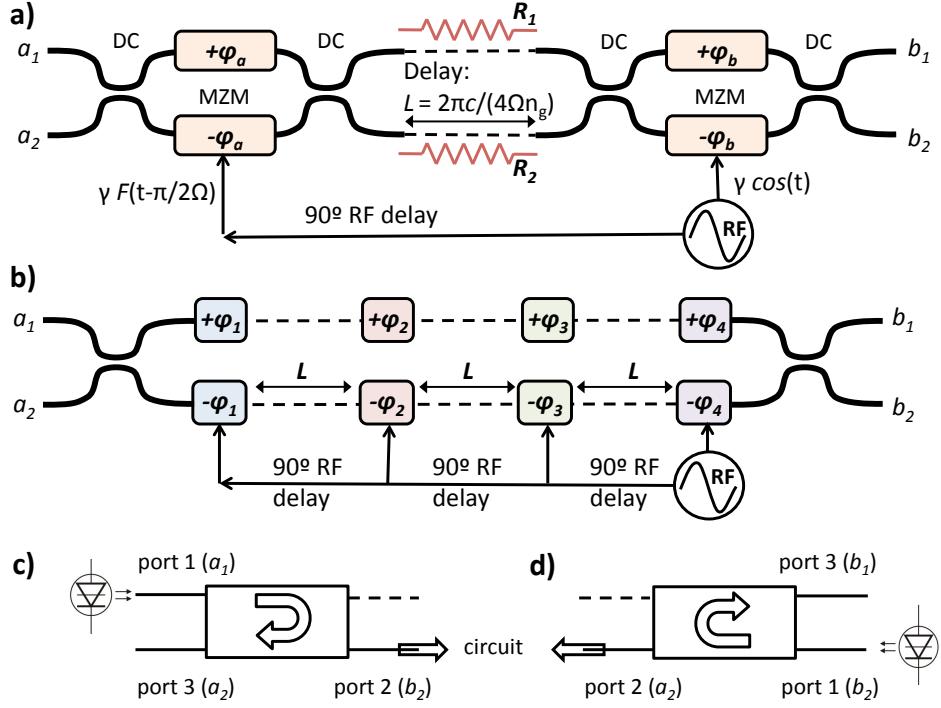

**Figure 4:** schematics of the two designs and their use as optical circulators. a) Design discussed in the main text (Fig. 1a). b) Alternative design discussed here. c) "Wiring" of design (a) to be used as a non-reciprocal modulator. d) Same for design (b). The two designs are equivalent after reversing the light propagation direction and switching the "thru" and "cross" ports.

Recalling the transfer matrices obtained for design (a):

$$\vec{T} = \begin{pmatrix} -\cos(2\gamma F(t)) & -\sin(2\gamma F(t)) \\ \sin(2\gamma F(t)) & -\cos(2\gamma F(t)) \end{pmatrix} \text{ and } \bar{T} = \begin{pmatrix} -1 & 0 \\ 0 & -1 \end{pmatrix}$$

We see that as far as the optical field intensity is concerned, the two designs perform the exact same function, after swapping both the roles of the "cross" and "thru" ports and the propagation direction. The two designs also have the same modulation efficiency and phase shift requirement, since the factor multiplying $\gamma$ scales with the number of phase modulators in each scheme. While the first design requires two additional beam splitters, it features three-times shorter optical and RF delay lines, yielding lower optical losses, electrical signal attenuation and electronics complexity. For the sake of simplicity, we performed our analysis in the main text based on the first design, but simulations where made with this alternative design yielding similar results.

In order to give an intuitive understanding of the second device functioning, we consider an ideal square-wave signal (no bandwidth limitation) and split each period into 4 equal intervals $\Delta t_i$, $i = 1..4$ (see Fig. 5). We can label any time interval with the convention $\Delta t_{i+4} = \Delta t_i$, $i$ any integer. We denote the phase shift applied on the upper (resp. lower) arm by modulator number $j$ at time $i$ by the matrix element $\varphi_{ij}$ (resp. $-\varphi_{ij}$). With this notation, light traveling from left to right accumulates a phase shift in the upper/lower arm of $\vec{\varphi}_{+/-} = \pm \sum_{k=0..3} \varphi_{i+k, j+k}$ (with the convention $\varphi_{i, j+4} = \varphi_{i,j}$). This corresponds to summing over the diagonals of the square matrix $\varphi_{ij}$ ($i,j=1..4$). For light traveling in "backward" direction (from right to left here), the accumulated phase shift writes: $\bar{\varphi}_{+/-} = \pm \sum_{k=0..3} \varphi_{i+k, j-k}$ , corresponding to a sum over the *anti*-diagonals. With this insight, it is readily seen that the following matrix

$$\varphi_{i,j} = \frac{\pi}{8} \begin{pmatrix} - & - & + & + \\ - & + & + & - \\ + & + & - & - \\ + & - & - & + \end{pmatrix}$$

always leads to a null phase shift in the forward direction, while backward propagating light experiences $\pm\pi/2$ optical phase shift per arm, corresponding to the condition for destructive interference in cross-arm transmission.

Using this matrix approach, it is easy to see that 4 modulation sections is the minimum number to achieve optical isolation.

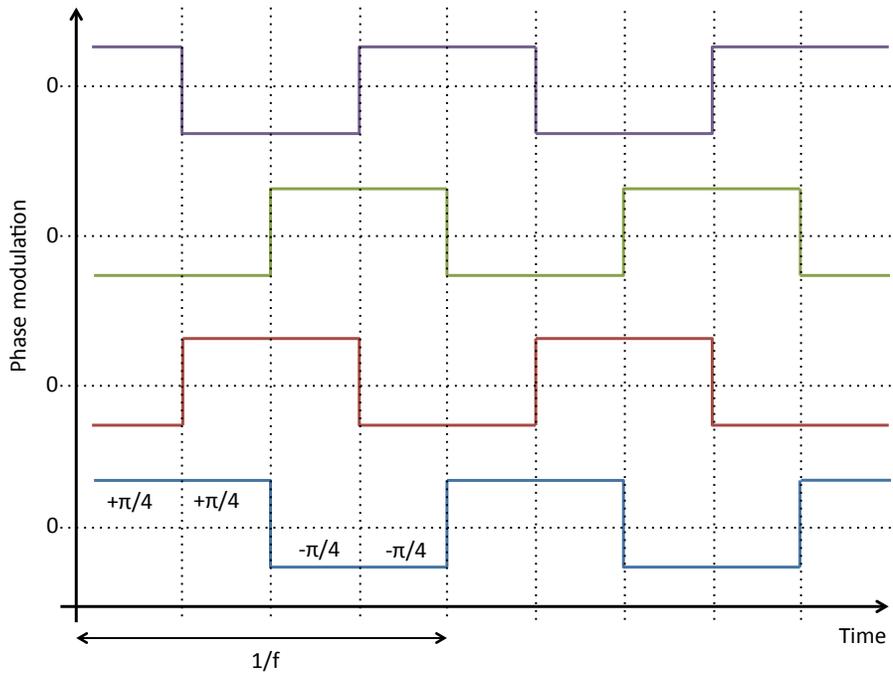

**Figure 5**: Driving 4 phase modulators with a retarded square-wave signal.